# Next-generation Co-Packaged Optics for Future Disaggregated AI Systems

Sajjad Moazeni
Electrical & Computer Engineering Department
University of Washington
Seattle, USA
smoazeni@uw.edu

*Abstract*—Co-packaged optics is poised to solve the interconnect bandwidth bottleneck for GPUs and AI accelerators in near future. This technology can immediately boost today's AI/ML compute power to train larger neural networks that can perform more complex tasks. More importantly, co-packaged optics unlocks new system-level opportunities to rethink our conventional supercomputing and datacenter architectures. Disaggregation of memory and compute units is one of such new paradigms that can greatly speed up AI/ML workloads by providing low-latency and high-throughput performance, while maintaining flexibility to support conventional cloud computing applications as well. This paper gives a brief overview of state-of-the-art of co-packaged optical I/O and requirements of its next generations. We also discuss ideas to exploit co-packaged optics in disaggregated AI systems and possible future directions.

## I. Introduction

Ever-increasing size and complexity of artificial intelligence (AI) and machine learning (ML) models and datasets have recently reached over a hundred trillion parameters [1]. This has created a vital need for data and model parallelization, distributed over thousands of processing and memory nodes. These nodes require low-latency and ultra-low power multi-Tb/s inter/intra-rack optical I/O as well as chip-to-chip interconnects. As an example, the latest Nvidia DGX H100 system for ML training consists of eight GPUs, each one needs a 7.2Tb/s bidirectional off-package bandwidth [2]. Recently, multiple interconnect technologies such as Compute Express Link (CXL) and NVLink have been emerged to answer this need and deliver high bandwidth, low-latency connectivity between processors, accelerators, network switches and controllers. Although currently these chip-to-chip links are realized via copper-based electrical links, they cannot meet the stringent speed, energy-efficiency, and bandwidth density requirements of future distributed AI computing. Silicon photonic transceivers have shown a great promise to address this challenge by ultimately co-packaging optical I/O with high performance CPUs and GPUs. Recent demonstration showed that we can utilize wavelength-division multiplexing (WDM) to boost aggregate bandwidth per fiber up to near 1Tb/s by transmitting data over multiple wavelengths in parallel [4]. Co-packaged optics will become an imminent part of network top-of-the-rack (ToR) switches and GPUs as recently announced by Broadcom and Nvidia. However, current demonstrations of co-packaged optics do not yet satisfy the needs of next-generation AI accelerators and ToR switches which require +10Tb/s bandwidths at sub-pJ/b energy-efficiency. We will further discuss the state-of-the-art and future performance metrics of optical I/O in Section II.

Co-packaged optics will also ignite revolutionary architecture-level paradigm shifts by providing ultrahigh bandwidths at low latencies that can reach up to a *km*-range distance. Disaggregated datacenter architecture is one of these possibilities that can decouple memory and storage (e.g., DRAM) from processors and accelerators (e.g., CPU, GPU) [3]. This scheme provides flexible and dynamic resource allocation for conventional diverse workloads in a datacenter. Recently, the disaggregation was also motivated by AI/ML computing as well to provide large amounts of memory access at reduced latency for processor nodes using features such as remote direct memory access (RDMA) and GPU-Direct. We will briefly discuss new ideas and future directions in Section III.

## II. State-of-the-art Co-packaged Optics

Today's high-performance compute nodes integrate processor dies, high-bandwidth memories (HBM), and soon co-packaged optics in a single 2.5D/3D package known as a system-in-package (SiP). The co-packaged optics block/chip comprised of silicon photonic optical transceiver dies with either on-chip or off-package WDM laser sources. Given the cost and limited number of available wavelengths (8, 16 or 32) in today's WDM sources, it is highly desired to achieve +100Gb/s/wavelength data-rates to increase bandwidth density and amortize the static laser power overhead. Here we discuss major challenges to achieve this target data-rate in terms of following key performance metrics:

### A. Energy-effiecny

Micro-ring modulators (MRM) have significantly improved the energy-efficiency of latest silicon photonics transmitters compared with traditional bulky Mach-Zehnder interferometer (MZI) modulators by an order of magnitude. The MRM-based transceivers recently achieved ~6pJ/b (including ~1pJ/b laser energy) energy-efficiency at 25Gb/s/wavelength over 8 WDM channels in a 45nm monolithic silicon photonic process [4]. Higher data-rates in this process leads to lower energy-efficiency due to the limitations of the 45nm CMOS. Using more advanced CMOS requires heterogenous electronic-photonic integration, which suffers from larger device parasitics. Either way, equalization is required for data-rates beyond +100Gb/s to overcome bandwidth limitations caused by limited CMOS speed, integrations (e.g., wire-bond or micro-bumps) parasitics, and photonic devices such as modulators. However, the equalizer circuitry can easily dominate the total link power even in the most advanced CMOS nodes. For example, in Intel's 112Gb/s optical transmitter [5], +25% of power is consumed by transmitter equalizer, leading to 6pJ/b energy-efficiency (only for the transmitter). To support +10Tb/s at reasonable power budgets for next generation co-packaged optics, sub-pJ/b energy-efficiencies will be required. Achieving this goal demands multiple technological breakthroughs since currently only the clocking and equalization circuitry each require +1pJ/b in a transceiver. Finally, the energy of in-package interconnect should be also considered which is currently 0.5-1pJ/b.

*B. Bandwidth Density*

Bandwidth density in terms of Tb/s per mm of package edge ("shoreline") is a critical metric which relies on multiple factors including: (1) Fiber packaging technology and density of fiber connectors into the package, (2) areal chip limits (area dedicated to the co-packaged optics and areal efficiency of a transceiver die in terms of Tb/s/mm$^2$), (3) ultimately bandwidth edge density of in-package I/O. Latest standard (UCIe) provides 1.3TB/s/mm (1mm edge of a chip) assuming advanced 45μm pitch. While today's co-packaged optics demos achieved ~0.2Tb/s/mm, densities of +1Tb/s/mm is expected for next generations.

*C. Latency*

Latency requirements mainly depends on the interconnect type (chip-to-chip, inter/intra-rack, etc.) and the computing application. If the optical link can achieve error-free (bit error rate (BER) <10$^{-12}$) communication without any forward-error correction (FEC) coding, the latencies are typically can be in the range of 10-100ns (limited by the time-of-flight and SerDes), which is acceptable for most of today's applications. However, this is very challenging at higher data-rates (+100Gb/s) at pJ/b-range energy-efficiency.

### III. CO-PACKGED OPTICS: MORE THAN JUST AN E/O BRIDGE

In the current generations, co-packaged optics is only envisioned to play a role of an electro-optical (E/O) bridge as shown in Fig. 1a to solve the interconnect bandwidth density bottleneck of SiPs. This, of course, can significantly improve overall performance (compute time and energy) for AI/ML algorithms in particular distributed training [6]. In addition, since co-packaged optics can facilitate datacenter disaggregation since the ultra-high bandwidth optical data can travel at much larger distances with only a minimal latency overhead. However, adding more functionalities to the co-packaged optics chip can enable novel architectures, where it can be "smarter" than being just an E/O bridge that has been envisioned so far. In doing so, the co-packaged optics can be treated as a "co-processor" in a SiP.

One such new functionality is to give HBM access to co-packaged optics as illustrated in Fig. 1b. In this architecture, co-packaged optics chip can access the memory without any intervention by the processor chip and it can potentially provide DMA with bypassing CPU/GPU buffering between any processor or DRAM SiPs. This provides a novel disaggregated AI architecture down to the package level without a need for restructuring or relocating the resources and pooling the memory and processors physically.

This new architecture can also greatly benefit multi-GPU nodes like Nvidia DGX machines by reducing the near memory access time closer to the local memory access latency (~5μs). The memory access latency been improved in today's systems by approaches such as GPU-Direct which bypasses unnecessary CPU buffering and using high bandwidth NVLinks between GPUs. Currently the latency to access neighboring GPU's HBM via a direct NVLink-V2 link is ~8μs [7]. Proposed architecture in Fig. 1b, can potentially reduce this latency by more than 25% with bypassing a large crossbar and L2-caching inside of the GPU. If we account for 1μs estimated overhead in this approach (due to the need of a crossbar and buffers on the co-packaged optics chip), any GPU can access neighboring GPU's HBM with ~6μs latency. This method can be extended beyond two neighboring GPUs using high-radix optical μs-switching networks [8].

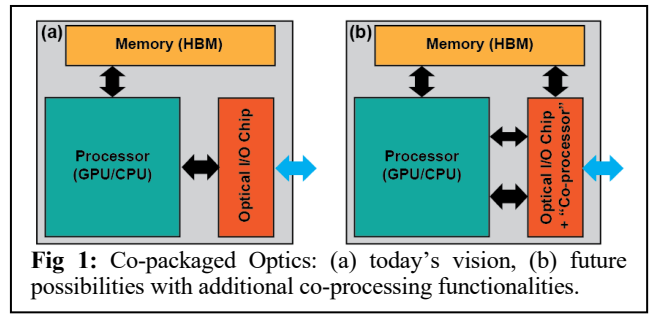

**Fig 1:** Co-packaged Optics: (a) today's vision, (b) future possibilities with additional co-processing functionalities.

Other possibilities are scenarios where the "smart" co-packaged optics can perform pre/post processing of data to reduce the communication bandwidth requirement and energy between the processer and co-packed optics chip. The advantages greatly depend on the compute workload and details of co-packaged optics implementations.

### IV. CONCLUSION

Current generations of co-packaged optics are about to revolutionize datacenter and supercomputing systems. Besides the challenges discussed in this paper, optical packaging, laser integration/cost, and thermal management should be considered. Next-generations of this technology requires about 5x-10x improvement in energy-efficiency and bandwidth density. From a system-level perspective, this technology brings in new possibilities including adding memory access and compute functionalities to utilize co-packaged optics as a "co-processor".

### ACKNOWLEDGMENT

This work is supported by NSF CAREER (ECCS-2142996).